\documentclass[twocolumn,showpacs,preprintnumbers,amsmath,amssymb,floatfix,aps,ajp]{revtex4}

\usepackage{latexsym,amssymb,amsfonts,amsthm,amscd}
\usepackage[dvipdfm]{graphicx}
\usepackage{comment}

\begin{document}
\newcommand{\n}{\~n}
\newcommand{\pderiv}[2]{\frac{\partial #1}{\partial #2}}
\newcommand{\tderiv}[2]{\frac{D#1}{D#2}}
\newcommand{\deriv}[2]{\frac{d#1}{d#2}}
\newcommand{\pd}[2]{\frac{\partial^2 #1}{\partial #2^2}}
\newcommand{\td}[2]{\frac{d^2#1}{d#2^2}}
\newcommand{\N}{\mathbb{N}}
\input{epsf}

\newif\ifcolor \colorfalse

\title{El Ni\n o and the Delayed Action Oscillator}
\author{Ian Boutle, Richard H.\ S.\ Taylor, Rudolf A.\ R\"{o}mer}
\affiliation{Department of Physics and Centre for Scientific
Computing, University of Warwick, Coventry CV4 7AL, United Kingdom}

\date{$Revision: 1.101 $, compiled \today}

\begin{abstract}
  We study the dynamics of the El Ni\~no phenomenon using the
  mathematical model of delayed-action oscillator (DAO). Topics such as
  the influence of the annual cycle, global warming, stochastic
  influences due to weather conditions and even off-equatorial
  heat-sinks can all be discussed using only modest analytical and
  numerical resources. Thus the DAO allows for a pedagogical
  introduction to the science of El Ni\~no and La Ni\~na while at the
  same time avoiding the need for large-scale computing resources
  normally associated with much more sophisticated coupled atmosphere-ocean
  general circulation models. It is an approach which is
  ideally suited for student projects both at high school and
  undergraduate level.
\end{abstract}
\maketitle

\section{Introduction}
\label{sec-introduction}

Where weather is concerned, attempts at its prediction
\cite{Bla43,SchZ99} and its apparent unpredictability have always
intrigued the human mind.  Related phenomena such as storms, floods
and droughts, as well as global weather and issues of climate
prediction such as the green-house-gas effect and global warming
regularly appear in news bulletins and TV programs worldwide
\cite{CNN97,Sup98}. Thus weather is where many of us get our first
exposure to science and in particular science's attempts at
predicting natural behavior of highly complex systems. However,
precisely because of this complexity, inexperienced science
enthusiasts rarely have the tools needed to engage in predicting
atmospheric behavior.

One particular example of a natural atmospheric phenomenon which
continues to attract regular public attention is the so-called El
Ni\~no event in the equatorial Pacific, which occurs irregularly at
intervals of $2$--$7$ years with occasionally dramatic consequences
for many regions worldwide. These ``teleconnections'' can be
observed in many places, for example spring rainfall levels in
Central Europe, flooding in East Africa, and the ferocity of the
hurricane season in the Gulf of Mexico \cite{Sau01}. Predictions of
when and where El Ni\~no weather patterns will dominate local
weather conditions are very hard to obtain. Usually, so-called
coupled atmosphere-ocean general circulation models (CGCMs),
combined with large scale computing resources are needed to study
and predict El Ni\~no's consequences \cite{PicMP97,KimL98}. Clearly,
such resources will normally be beyond the reach of teachers and
even lecturers at university. Consequently, student projects on the
El Ni\~no phenomenon which involve a substantial {\em quantitative}
component are rare.

Fortunately, a route towards qualitative {\em and} quantitative
prediction exists. This approach is based on the {\em delayed-action
oscillator} (DAO), a one-dimensional, non-linear ordinary
differential equation with delay term which was introduced in 1988
by Suarez and Schopf \cite{SuaS88}. The DAO models the irregular
fluctuations of the sea-surface temperature (SST), incorporating the
full coupled Navier-Stokes dynamics of the El Ni\~no event via a
suitably chosen non-linearity. While previous studies of the model
have concentrated on the universal properties of the dimensionless
dynamics \cite{SuaS88,WhiTBD03}, we shall show in the present paper
how to use the DAO to predict many of the essential features of El
Ni\~no events. We shall extend the model to quantitatively include
the effects of the annual cycle, global warming, randomness in
temperature conditions and the possible influence of the dynamics
outside the equator.
We believe that this discussion of the DAO opens up various avenues
for student projects, be it at the high school or undergraduate
level. Important underlying physics and meteorology concepts such as
ocean waves and their speeds, non-linear dynamics, the Coriolis
force, convective cycles of ocean and atmosphere currents, global
wind patterns and their stability can be discussed and help the
students understand that El Ni\~no --- far from only being a bringer
of doom as sometimes portrayed in the media --- is in fact a natural
consequence of our planet's celestial dynamics \cite{BurS98}.

The paper is organized as follows. In Section \ref{sec-ENSO}, we
review the El Ni\~no-Southern Oscillation phenomenon. Section
\ref{sec-DAO} introduces the basic DAO model and we study its basic
dynamics. Sections \ref{sec-annualforcing}, \ref{sec-globalwarming}
and \ref{sec-randomeffects} investigate the influences of annual
forcing, global warming and random effects on the DAO, with Section
\ref{sec-allinone} incorporating all these processes. Section
\ref{sec-coupledDAOs} then extends that single DAO to more than one
coupled DAO. We conclude in Section \ref{sec-conclusions}.
Information on the numerical methods and an example code can be
found in the appendix.

\section{ENSO Phenomenology}
\label{sec-ENSO}

The term El Ni\~no was originally used, from as early as the late
19th century \cite{Phi90}, in reference to an observed warm
southward flowing current that moderates the low SST along the west
coast of Peru and Ecuador \cite{NOAA,COAPS}. This current arrives
after Christmas, during the early months of the calendar year, and
was therefore named ``El Ni\~no'' (The ``Little Boy'' or Christ
Child in Spanish). In certain years this current would be unusually
strong, bringing heavy rains and flooding inland, but also
decimating fishing stocks, bird populations and other water-based
wildlife in what would normally be an abundant part of the Pacific.

\begin{figure}[tbhp]
  \center
  \includegraphics[width=0.9\columnwidth]{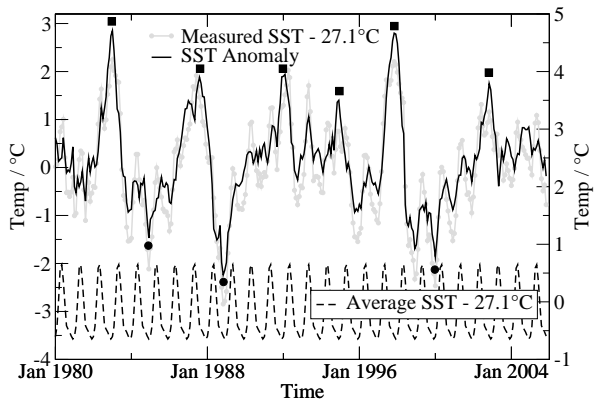}
  \caption{ Observed ENSO SST oscillations since $1980$. The average
    (dashed black, right axis) and measured (grey, left axis) SST's,
    minus $27.1^\circ$C, in the Ni{\n}o $3.4$ region (cp.\ Fig.\ \ref{fig-ENSO-geometry}) with the SST
    anomaly (solid black, left axis). El Ni{\n}o events are denoted by a
    black square above the event, La Ni{\n}a events (see section \ref{sec-lanina}) by a black circle below.}
\label{fig-SSToscillations}
\end{figure}

Today, the term El Ni\~no is most often used when describing a far
larger-scale warm event that can be observed across the whole of the
Pacific Ocean by certain characteristic climatic conditions. El
Ni\~no is just one phase of the so-called El Ni\~no-Southern
Oscillation (ENSO) phenomenon, i.e.\ an irregular cycle of coupled
ocean temperature and atmospheric pressure oscillations across the
whole equatorial Pacific. In Fig.\ \ref{fig-SSToscillations}, we
show the measured SST as well as the average annual cycle.

Since 1899, there have been $29$ recorded El Ni\~no events, giving
an average period for ENSO of $3.7$ years \cite{KovBHW03}. However,
there is evidence to suggest that El Ni\~no events are becoming
stronger and more frequent; there have been $6$ events in the last
$20$ yrs (an average period of $3.3$ years), $2$ of which were the
strongest of the century \cite{KovBHW03,NOAA}.

\subsection{Normal conditions}
\label{sec-normalconditions}

\begin{figure}[tbhp]
  \center
  \includegraphics[width=0.9\columnwidth]{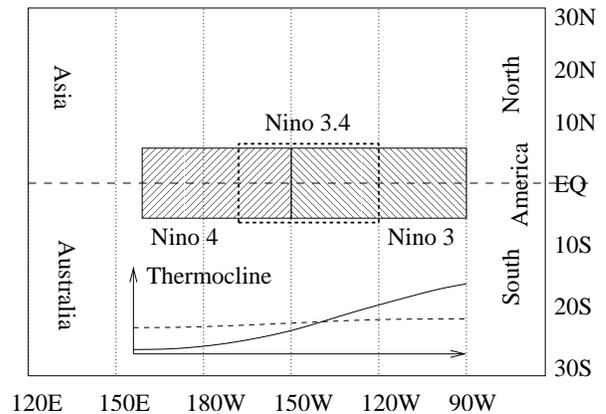}
  \caption{
  Schematic view of the Ni\~no regions $3$ (right hashed rectangle),
  $4$ (left hashed rectangle) and $3.4$ (rectangle with dashed line borders).
  The thermocline height in normal (solid) and El Ni\~no conditions (dashed line) is
  indicated schematically in the diagram.}
\label{fig-ENSO-geometry}
\end{figure}

In order to understand the El Ni\~no anomaly, let us first describe
the normal conditions that exist in the Pacific Ocean between
$120^\circ$ East and $90^\circ$ West. In Fig.\
\ref{fig-ENSO-geometry} we show a schematic view of the relevant
section of the Pacific.

In normal conditions, the trade winds blow from East to West across
the tropical Pacific, resulting in warm surface water being pushed
west, so that the thermocline --- a region of water roughly $100$m
below the surface where there is a very sharp temperature gradient
separating the warm surface layer from the cold abyss --- is
depressed further in the West, and made shallower in the East
\cite{Phi90}. The SST is about $8^\circ$C higher in the West, with
cool temperatures off South America, due to cold water {upwelling},
i.e.\ rising from the deeper layer to the surface \cite{Mad00}. This
cold nutrient-rich water supports diverse marine ecosystems and
major fisheries, so upwelling is of immense ecological and economic
importance to the west coast of the Americas \cite{NOAA,COAPS}.

The warmer SST in the West means increased evaporation and rising
moist air, and hence high levels of rainfall (a tropical climate),
along with low atmospheric sea-level pressures in the West. In the
East, the colder SST results in far less rainfall, an arid inland
climate and higher atmospheric sea-level pressures.
This induced pressure difference drives the circulation of the trade winds. A
measure of the strength of the trade winds is the Southern Oscillation Index
(SOI), defined as the normalized difference in sea-level pressures between Tahiti
and Darwin, Australia.

The whole process is a therefore a self-sustaining cycle; winds
cause thermocline and SST changes, which cause pressure differences,
which in turn drive the winds.

\subsection{El Ni\~no conditions}
\label{sec-elnino}

During an El Ni\~no event the SST and SOI change dramatically.
In the central and western Pacific, there is relaxation in the
strength of the trade winds. This leads to a depression of the
thermocline in the East and an elevation in the West, because less
warm water is being pushed west. This increase in thermocline depth
in the East can be as large as a factor of three, leading to an
increase in SST, and the effective cut-off of upwelling of nutrients
to the surface. This has devastating consequences for the primary
producers (such as plankton) in the food chain, with knock-on
effects for other life, such as fish and birdlife \cite{Phi90}.

The increased SST in the East also leads to an increased
evaporation, and large amounts of rain, often causing extensive
flooding. The rising air decreases the normally high pressure.
During large El Ni\~no events the SOI is very negative, indicating a
reversal in trade winds. This causes a great deal of warm, moist air
(and hence rain) to move east, whilst leaving a severe drought in
the western Pacific, often resulting in large bush fires.

\subsection{La Ni\~na conditions}
\label{sec-lanina}

La Ni\~na (The ``Little Girl'') is the phase of the ENSO cycle which is
opposite to El Ni\~no. It is characterized by unusually {\em cold}
ocean temperatures in the central and eastern Pacific. La Ni\~na arises
by a strengthening in the trade winds, further depressing the
thermocline in the West (and elevating it in the East), which increases
cold water upwelling in the eastern Pacific. This lower SST means the
high levels of nutrients can support an abundance of life in the Ocean.
The SOI is also positive (increased pressure difference), bringing
enhanced rainfall in the West.

\section{The Delayed-Action Oscillator}
\label{sec-DAO}

\subsection{A non-linear oscillator with delay term}

The DAO is a simple non-linear model in which the ENSO effect is
modeled by just three terms \cite{SuaS88}. With $T$ denoting the
temperature {\em anomaly}, i.e.\ the deviation from a suitably
defined long term average temperature, the model is given as a
first-order, non-linear differential equation
\begin{equation} \label{eq-DAO}
\frac{dT}{dt}=\ k T -\ b T^3 -\ {\cal A} T(t-\Delta)
\end{equation}
with appropriate coupling constants $k$, $b$, ${\cal A}$ and
$\Delta$.
The first term on the right hand side in (\ref{eq-DAO}) represents a
strong {\em positive} feedback within a coupled ocean-atmosphere
system which is assumed to occur in the Ni{\n}o $3.4$ region. A much
deeper thermocline as in the West, or a cold SST as in the East,
would result in much poorer feedback coupling.
The feedback process works through advective processes giving rise
to temperature perturbations which in turn result in atmospheric
heating. This heating leads to surface winds, driving the ocean
currents which will then enhance the anomaly $T$.

The second component is an unspecified nonlinear net {\em damping} term
that is present in order to limit the growth of unstable perturbations.
This term represents various ocean advective processes and moist
processes that stop temperatures diverging.

Lastly, the model also considers {\em equatorially-trapped} ocean
waves propagating across the Pacific, before interacting back with
the central Pacific region after a certain time delay. These ocean
waves are ``hidden'' Rossby waves which move westward on the
thermocline, reflect off the rigid continental boundary in the West
and then return eastward along the equator as Kelvin waves. Their
respective wave velocities and the width of the Pacific basin
determine the transit time, which is denoted $\Delta$ in Eq.\
(\ref{eq-DAO}). The strength of the returning, emerging signals
relative to that of the local non-delayed feedback is denoted by
${\cal A}$. We also assume that $0<{\cal A}<k$ due to loss of
information during transit and imperfect reflection.

We shall often use the DAO in its dimensionless form
\begin{equation} \label{eq-DAOdimless}
\frac{dT'}{dt'}=\ T' -\ (T')^3 -\ \alpha T'(t'-\delta),
\end{equation}
where we have scaled time by $t'=k t$ and temperature by $T'=
\sqrt{b/k} T$. The new constants then are $\alpha= {\cal A}/k$ and
$\delta= k \Delta$. In what follows, we shall for simplicity
suppress the prime in all expressions where there is no ambiguity.

\subsection{DAO wave propagation}

The DAO delay term has a negative coefficient representing a
negative feedback. To see the reason for this, let us consider a
warm SST perturbation in the coupled region. This produces a
westerly wind response that deepens the thermocline locally
(immediate positive feedback), but at the same time, the wind
perturbations produce \emph{divergent} westward propagating Rossby
waves that return after time $\Delta$ to create upwelling and
cooling, reducing the original perturbation. This process can
therefore be seen as a phase-reversing reflection off the coupled
region; the resultant upwelling will cause further Rossby wave
propagation, which in turn will cause downwelling and warming after
a further delay, $\Delta$. Hence we expect the period of model
oscillations to be no less that $2\Delta$
--- ``period doubling'' \cite{SuaS88}.

Note that the eastern boundary will also contribute a delay term to
the DAO, however equatorial Kelvin wave reflection from eastern
boundaries is weak as the bulk of the wave propagates {\em along}
the coast \cite{SoaWW99}. For this reason we may ignore these
effects.

It is possible to estimate the wave transit times, and hence the
time delay $\Delta$ of the DAO model, from knowledge of the wave
speeds for Rossby and Kelvin waves (see Appendix
\ref{sec-shallowwaterwaves}). We shall return to this direct
comparison later, but for now we focus on the behavior of the
dimensionless DAO (\ref{eq-DAOdimless}).

\subsection{Linear stability analysis}
\label{sec-DAO-linearstability}

Equation (\ref{eq-DAOdimless}) cannot be solved analytically. But we
may study its dynamics by analyzing its behavior close the its fixed
points, i.e.\ those special solutions $\bar{T}$ of
(\ref{eq-DAOdimless}) for which ${dT}/{dt}=0$ which also immediately
implies $T(t)=T(t-\delta)$ for all $t$. Hence we need to solve
\begin{equation}
(1-\alpha) \bar{T} -\ {\bar{T}}^3 = 0
\end{equation}
and so the fixed point solutions are $\bar{T} = 0, \pm
\sqrt{1-\alpha}$. $\bar{T}_0=0$ is the trivial fixed point,
corresponding to a zero temperature anomaly. It is also unstable,
i.e.\ any small perturbation will lead to a system moving away from
$0$. Therefore we shall concentrate on the other two fixed points
$\bar{T}_\pm=\pm\sqrt{1-\alpha}$. Due to symmetry, it is furthermore
sufficient to consider one of these only and we choose
$\bar{T}_+=\sqrt{1-\alpha}$.

The dynamics close to $\bar{T}_+$ can be examined by a linear
stability analysis. Let us consider a small perturbation $S = T -
\bar{T}_+$. Ignoring terms of more than linear order in $S$
\cite{SuaS88}, we find when substituting $S$ back into
$(\ref{eq-DAOdimless})$ that
\begin{equation} \label{eq-linearDAO}
\frac{dS}{dt} =\ (3\alpha -2) S -\ \alpha S(t- \delta )
\end{equation}
We now look for solutions of the form $S = A e^{\sigma t}$, where
$A$ is a constant real number equal the initial condition $S(0)$ and
$\sigma$ is a complex number. Substituting this in
(\ref{eq-linearDAO}) gives an equation for $\sigma$
\begin{equation}\label{eq-sigma}
    \sigma = (3\alpha -2) -\ \alpha e^{-\sigma \delta}
\end{equation}
Writing $\sigma=\sigma _{\rm R} + \imath \sigma _{\rm I}$, where $\sigma _{\rm R}, \sigma _{\rm I}$ are both real, we find
\begin{subequations}
\begin{eqnarray}
 \sigma_{\rm R} &= &(3\alpha -2) -\ \alpha \ \cos(\sigma_{\rm I} \delta) \ e^{-\sigma_{\rm R} \delta} \\
  \sigma_{\rm I} &= &\alpha \ \sin(\sigma_{\rm I} \delta) \ e^{-\sigma_{\rm R} \delta}
\end{eqnarray}
\end{subequations}
When $\sigma_{\rm R}=0$, the solution $S = A e^{\imath \sigma_{\rm
I} t}$ is purely oscillatory and hence the fixed point is a {\em
center} --- neither stable nor unstable --- with a finite frequency
of $2 \pi / \sigma_{\rm I}$. These {\em neutral} solutions occur
along (an infinite number of) neutral curves of the form:
\begin{equation}\label{eq-neutralcurve}
    \delta = \frac{2n \pi \mp \arccos\left(\frac{3 \alpha - 2}{\alpha}\right)
}{ \sqrt{ \alpha^2- \left(3\alpha - 2 \right)^2 }}\  , \qquad n \in
\mathbb{N} \cup \{0\}
\end{equation}

\begin{figure}[tbhp]
\includegraphics[width=0.9\columnwidth]{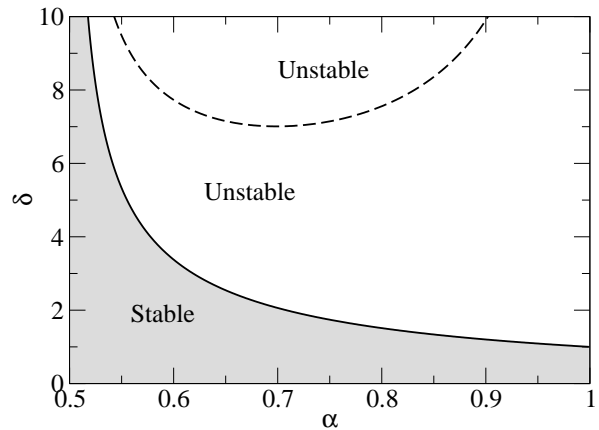}
\caption{
    Phase diagram for the stability of fixed points $\bar{T}_+$. Neutral
    curves for $n=0$ (solid line) and $n=1$ (dashed line) are shown
    together with stable (grey) and unstable regions (white).
    }
\label{fig-neutralcurves}
\end{figure}

In Figure \ref{fig-neutralcurves}, we show the $(\alpha, \delta)$
phase diagram of the DAO obtained from the linear stability
analysis.
Only $0.5<\alpha<1$ is plotted, since for this domain the
coefficient on the local term in $(\ref{eq-linearDAO})$ is always
smaller than that of the delayed term. If it were larger, then the
local term would always dominate, and the solution would always be
stable, i.e.\ for $\alpha<0.5$ the fixed point is always stable and
attracting, regardless of the initial conditions.

Choosing ($\alpha,\delta$) below the first neutral curve, e.g.\
within the shaded region of Figure \ref{fig-neutralcurves},
corresponds to $\sigma_{\rm R}<0$ and hence the solution is a
decaying spiral
--- the fixed point is stable and attractive. However, taking
($\alpha$, $\delta$) above this neutral curve gives $\sigma_{\rm
R}>0$, and so the perturbation spiral grows outwards --- the fixed
point is unstable and no matter how close to $T_0$ the solution
starts, it will always move away.

If ($\alpha,\delta$) is taken above the second neutral curve, the
fixed point is still unstable $(\sigma_{\rm R}>0)$ but the number of
unstable modes has increased to two. However, the basic DAO equation
(\ref{eq-DAOdimless}) is only a one-dimensional problem, as there is
only one variable $T$, and so the extra unstable modes do not
influence the dynamics.

\begin{figure}[tbhp]
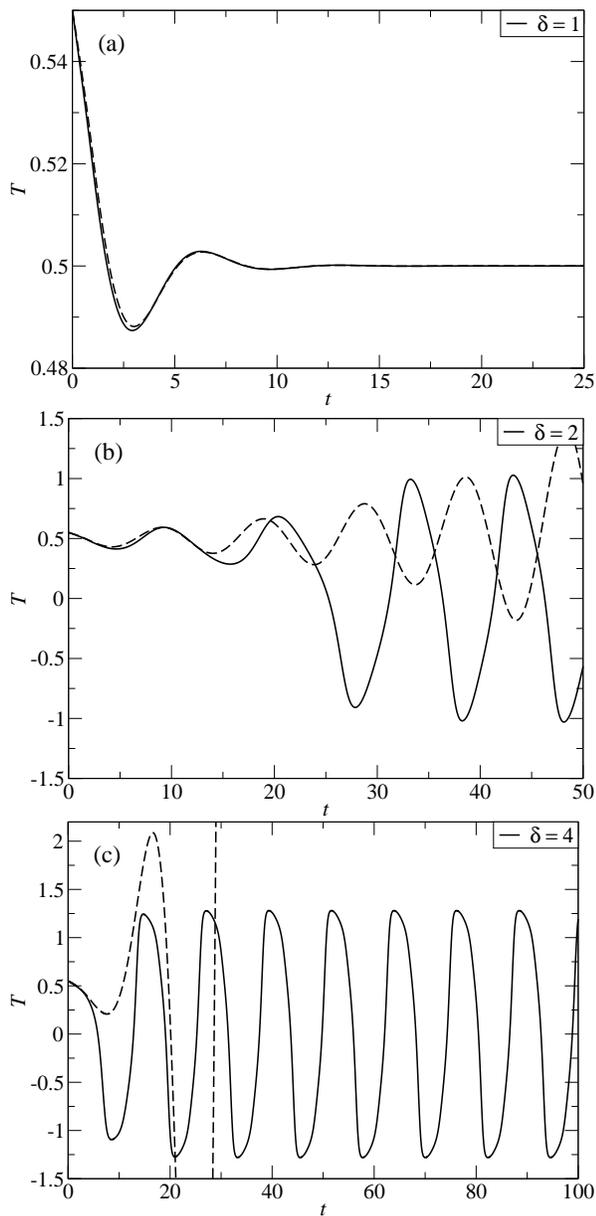

\centerline{
 \includegraphics[width=0.9\columnwidth]{Images/fig-basicsolutions_a.eps}}
\centerline{
 \includegraphics[width=0.9\columnwidth]{Images/fig-basicsolutions_b.eps}}
\centerline{
 \includegraphics[width=0.9\columnwidth]{Images/fig-basicsolutions_c.eps}}
\caption{An $\alpha=0.75$ cross-section for $T$ vs $t$ for (a) a stable
  solution with $\delta=1$, (b) a nearly neutral solution at
  $\delta=2$ and (c) an unstable oscillation at $\delta=4$. Solid
  lines indicate numerical solutions for the anomaly $T$, dashed lines
  are obtained from the linear stability solutions $\bar{T}_+ + S$. For clarity, we only plot the $\delta=4$ linear stability solution up to $t=30$.}
\label{fig-basicsolutions}
\end{figure}

This linear stability approach is highly useful, given that it
depends on only two parameters, $\alpha$ and $\delta$, and reveals
valuable information about the stabilities of the fixed points.
Given an $\alpha$ and a $\delta$, it also provides a solution to the
DAO given by:
\begin{equation} \label{eq-linearsolution}
T(t) = \bar{T}_+ + A e^{\sigma _{\rm R} t} e^{\imath \sigma _{\rm I}
t}
\end{equation}
which is perfectly valid to within an order of ${\cal O}(T)$,
provided it stays close to $\bar{T}_+$. However, once a solution is
too far from an unstable fixed point or if the initial condition
starts too far away from the fixed point, this linear stability
approach breaks down. Figure \ref{fig-basicsolutions} shows linear
stability solutions for different $\delta$ values, given by Eq. (\ref{eq-linearsolution}), along
with the full solutions discussed in Section
\ref{sec-fullsolutions}.

\subsection{Computation of full solutions by numerical integration}
\label{sec-fullsolutions}

Let us now construct solutions to the DAO which are outside the
region of validity of the linear stability analysis. As mentioned
before, this is no longer possible analytically, so we will use
numerical techniques to integrate up the time-behavior of
(\ref{eq-DAOdimless}). Fortunately, this involves the use of
standard numerical tools available in a variety of mathematical
software packages. For convenience, we have chosen the predefined
{\sl Mathematica} package \verb+NDelayDSolve+ \cite{Hay04}. The {\sl
Mathematica} code used to solve the basic DAO equation
(\ref{eq-DAOdimless}) is part of the code shown in Appendix
\ref{sec-numericalroutines}.

\begin{figure}[tbhp]
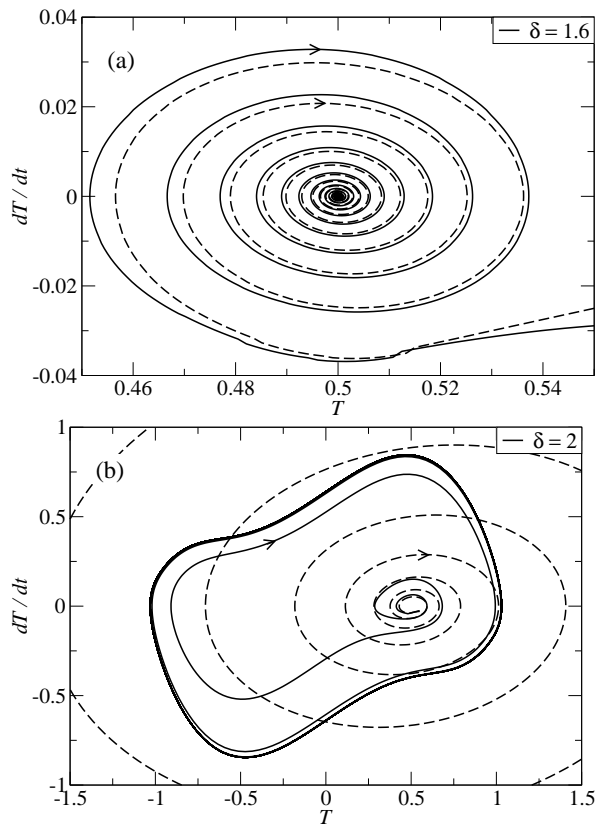

\centerline{
\includegraphics[width=0.9\columnwidth]{Images/fig-phaseportrait_a.eps}}
\centerline{
\includegraphics[width=0.9\columnwidth]{Images/fig-phaseportrait_b.eps}}
\caption{
    Phase portraits of $\frac{dT}{dt}$ vs $T$ with $\alpha=0.75$, solved
    up to $t_{\rm final}= 100\delta$ with (a) $\delta=1.6$ and (b)
    $\delta=2$. Solid lines indicate numerical solutions of Eq.\
    (\ref{eq-DAOdimless}), dashed lines correspond to approximations obtained
    from the linear stability analysis. The arrows indicate the time flow.}
\label{fig-phaseportrait}
\end{figure}

In Figure \ref{fig-basicsolutions}, we show the results obtained from Eq.\
(\ref{eq-DAOdimless}) when carrying out this numerical procedure for
$\alpha=0.75$ and various values of $\delta$. The initial value used is $T(0)=
0.55$ which is close to the fixed point $\bar{T}_+=0.5$. In Fig.\
\ref{fig-phaseportrait}, we show the $(T,dT/dt)$ phase portraits for two values
of $\delta$ close to the first neutral curve.

Let us now briefly discuss the observed behavior. In Fig.\
\ref{fig-basicsolutions}(a), we see that the numerical solution and
the linear approximation are in good agreement and quickly converge
to the stable fixed point $\bar{T}_+=0.5$ as expected for $\delta
=1$. In Fig.\ \ref{fig-basicsolutions}(b), we are still close to the
neutral curve but have already crossed over into the unstable
regime. Consequently, the numerical solution starts to slowly spiral
away from $\bar{T}_+=0.5$, becoming large enough so that the
dynamics are also influenced by the second unstable fixed point at
$\bar{T}_-= -0.5$, resulting in the onset of oscillations between
them. This is shown in more detail in Fig.\
\ref{fig-phaseportrait}(b). The agreement with the numerical
solution clearly breaks down very quickly, since the linear
stability analysis shows an unstable spiral, which continues to only
grow about $\bar{T}_+$.

At an even larger $\delta$ value, we see in Fig.\
\ref{fig-basicsolutions}(c) that the numerical solution immediately
diverges from the fixed point and goes to an oscillatory state, the
amplitude of which is limited by the nonlinear damping term in Eq.\
(\ref{eq-DAOdimless}).

The behavior for other values of $\alpha$ is analogous. Therefore,
we have identified within the DAO model (\ref{eq-DAOdimless}) a
whole region of $\alpha$ and $\delta$ values where we obtain
oscillations of the SST anomaly between a hot ($\bar{T}_+>0$) and a
cold ($\bar{T}_-<0$) fixed point. This feature is what makes the DAO
a model for oscillations between hot El Ni\~no and cold La Ni\~na
conditions.

\subsection{Comparing periodicities}

As identified in the last section, the DAO model is capable of
producing limit cycle oscillations for a wide region of ($\alpha,
\delta$) values. We next compute the period of these oscillations and
compare them to those observed for typical ENSO cycles.

As shown in Figure \ref{fig-ENSO-geometry}, the Ni\n o $3.4$ region
extends from $120^\circ$ West to $170^\circ$ West, with the midpoint
at $145^\circ$ West. If we assume that the Pacific western boundary
is at $\approx 120^\circ$ East, this gives an angular separation of
$95^\circ$ of longitude for the waves to travel. This corresponds to
a distance of $95 (2 \pi/360) r_{\rm Earth} = 10.6 \times 10^6$m,
where the radius of the Earth $r_{\rm Earth}=6.37\times 10^6$m.
Using the values of $1.4$ms$^{-1}$ for a Kelvin wave and
$0.47$ms$^{-1}$ for the primary Rossby mode as obtained in Appendix
\ref{sec-shallowwaterwaves} thus gives a delay of $262$ days for the
Rossby propagation to the western boundary, and a further $87$ days
for the return of the Kelvin waves, a total delay of $\Delta=349$
days.

\begin{figure}[tbhp]
  \ifcolor
  \centerline{\includegraphics[width=0.9\columnwidth]{Images/fig-DAOperiods-color.eps}}
  \else
  \centerline{\includegraphics[width=0.9\columnwidth]{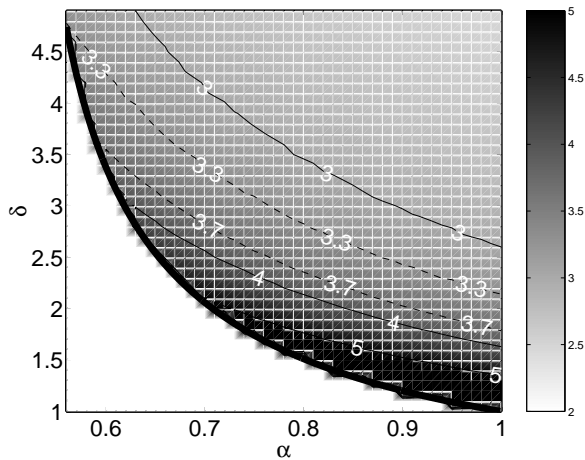}}
  \fi
  \caption{Grey scale plot of DAO periods for various $\alpha$ and $\delta$ values.
  Lines of constant periods are indicated, those corresponding to periods $3.3$ and $3.7$ are
  shown as dashed lines. The white region corresponds to the stable region as present in Fig.\ \ref{fig-neutralcurves},
  with the thick black line representing the first neutral curve.}
\label{fig-DAOperiods}
\end{figure}

This information now allows us to compute the time scaling
parameter, however since $k= \delta/\Delta$, these are different for
each value of $\delta$ used in the DAO model. By the use of a
Fourier Transform applied over $600$ years, we find e.g.\ for
$\alpha=0.7, \delta=3$ a period $\tau'=11.1$ which corresponds to a
true DAO period of $\tau= \tau' /k = 3.5$ years. In Fig.\
\ref{fig-DAOperiods} we show the results of many runs with varying
$\alpha= 0.56, 0.57, \ldots, 1$ and $\delta= 1, 1.1, \ldots, 4.9$.
Comparing these periods to real ENSO events, which have a period of
$3.3$ to  $3.7$ years, we see that there is a large region of
$(\alpha, \delta)$ values where there is a good agreement.

\subsection{Discussion of the basic DAO model}
\label{sec-DAO-Discussion}

In the preceding sections, we have seen that the DAO is a useful
tool in studying the ENSO cycle.
The main beauty of the DAO (\ref{eq-DAO}) is its simplicity.
Although the full equation cannot be solved explicitly, the linear
stability analysis and the numerical solutions both give highly
useful insights. In particular, the numerical solutions exhibit the
true global dynamics of the DAO. There are limit cycle oscillations,
with a period that matches the real world, and hence the DAO is a
good model for these features of ENSO. Nevertheless, the DAO model
assumes a localized coupled problem, which is obviously an
over-simplification of the situation in the Pacific. All influences
of the atmosphere, the earth's rotation, and the ocean currents and
dynamics are simply modeled by a non-linear, effective damping term
and a somewhat arbitrary delay. The delay term ignores, e.g.\ East
boundary reflections that could possibly be considered. The model
further assumes that time scales must be related to propagation
times of equatorially trapped oceanic waves in a closed basin. For
the atmosphere, which has short time scales, the DAO simply assumes
an {\em instantaneous} response.

\section{Annual Forcing}
\label{sec-annualforcing}

In this and the following sections, we shall extend the basic DAO
discussed above by including external influences which allow us to
model the ENSO dynamics better. In order to distinguish this
extended DAO from its original, we shall denote the new model as
ENSO-DAO from now on.

\subsection{Modifying the DAO to include the annual cycle}

As mentioned above, an El Ni{\n}o event typically appears in late
December, early January. Thus the annual cycle is clearly playing an
important role in determining the onset of El Ni{\n}o
\cite{ZebC87,Str99}. Let us thus add to the DAO (\ref{eq-DAO})
another term that models the periodic heating and cooling of the
equator during the annual cycle. The average SST data, as shown on
Fig.\ \ref{fig-SSToscillations}, can be used to construct a {\sl
Mathematica} interpolating function, $Y(t)$, representing the
continuous changes of the average SST during the year \footnote{We
remark that a simple sinusoidal term will also suffice for most
purposes.}. We therefore include its numerical time derivative,
representing the expected change in SST due to annual forcing, to
give a new DAO equation
\begin{equation} \label{eq-annualDAO}
\frac{dT}{dt}=\ k T -\ b T^3 -\ {\cal A}
T(t-\Delta)+\frac{d}{dt}Y(t)
\end{equation}
This equation now depends critically on the $(\alpha, \delta)$
values chosen in the dimensionless model, since computation of the
scaling constants $k$ and $b$ is required. We have seen earlier that
$k= \delta/\Delta$, and $b$ is given by the expression
$b=k(T'/T)^2$. In order to estimate $b$ we shall take $T'/T$ as the
ratio of their respective maxima/minima. Figure
\ref{fig-SSToscillations} shows the maximum anomaly is on average
$2^\circ$C, whilst the maximum value of $T'$ can be calculated for
any values $(\alpha, \delta)$.

Choosing $\alpha=0.7, \delta=3$ as the standard ENSO-DAO model for
the next few sections allows us to calculate $k=3.14$/years,
$b=1.09$ $^\circ$C$^{-2}$/years, and hence discuss the behavior of
this specific model under annual forcing.

\subsection{Solving the annually forced DAO}

Fixed points of (\ref{eq-annualDAO}) do not exist; they are now
fixed cycles fluctuating about $0, \pm \sqrt{(k-{\cal A})/b}$ due to
the periodic nature of the forcing term.
Nevertheless, taking $t=0$ to be the start of the year, the expected
SST will be at its coldest and $dY/dt=0$, so we can take the initial
condition to be $T(0) = \sqrt{(k-{\cal A})/b} + 0.15$ (this is
similar to before but now includes the scaling constants).

\begin{figure}[tbhp]
\includegraphics[width=0.9\columnwidth]{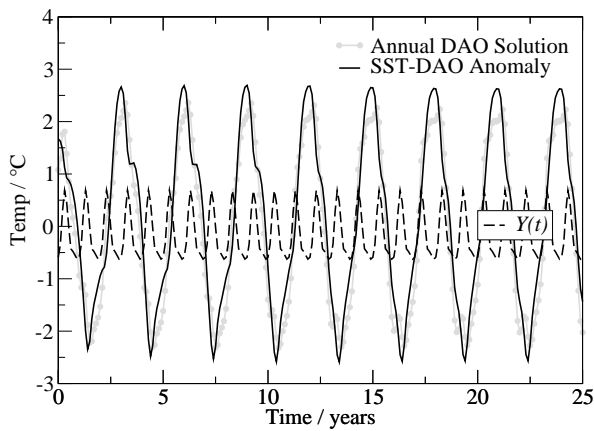}
\caption{
    Numerical solution of the annually forced DAO
    (\ref{eq-annualDAO}) with $\delta=3$ and $\alpha=0.7$ showing the annual cycle $Y(t)$ (dashed
    black), model solution $T$ (grey) and SST anomaly $T-Y(t)$ (solid
    black).}
\label{fig-annualDAO}
\end{figure}

Let us now consider numerical solutions of Eq.\ (\ref{eq-annualDAO}).
In Fig.\ \ref{fig-annualDAO}, we show the solution of $T$ vs $t$ for
parameters as above. The annual cycle as in Fig.\
\ref{fig-SSToscillations} is also shown.
We note that with the inclusion of annual forcing, $T$ is no longer
the SST {\em anomaly}, but rather, the anomaly is the {\em
difference} between $T$ and the annual forcing, $Y(t)$. This true
SST anomaly is also shown in Fig.\ \ref{fig-annualDAO}, and should
be compared to Fig.\ \ref{fig-SSToscillations}.

From Fig.\ \ref{fig-annualDAO} it is clear that the addition of an
annual forcing term has changed the dynamics in several ways.
Firstly, the peak amplitude of the oscillations has been increased.
The peaks themselves are less regular because of the quickly
changing forcing term provided by the seasons. However, the
paramount effect of the annual cycle is in determining the onset and
period of an El Ni\~no event. Hot events only occur \emph{in-phase}
with the annual cycle, and hence the ENSO-DAO now has an exact
integer period, e.g.\ 3 years for our model.
The ENSO-DAO gives high temperatures from July through to March of
the next year --- which is indeed as observed in the Pacific ---
with the maximum amplitude usually in December.

\section{Global Warming}
\label{sec-globalwarming}

Global warming is another effect that may influence the periodicity
and amplitude of ENSO events. For the purpose of this paper it will
be modeled via a constant heating on the Pacific DAO system. Working
with the dimensionless DAO for simplicity, this is achieved by
adding a constant $\beta$ to the DAO equation (\ref{eq-DAOdimless})
giving:
\begin{equation} \label{eq-forcedDAO}
\frac{dT}{dt}=\ T -\ T^3 -\ \alpha T(t-\delta) +\ \beta
\end{equation}
On its own, this term would give $\frac{dT}{dt} = \beta$, solved by
$T=\beta t +c$, a linear warming with time. We shall consider
$\beta>0$, a constant heating, although $\beta<0$ cooling can be
treated completely analogously.

Current climate change models predict the rise in average global
temperature over the next $100$ years will be below $5^\circ$C
\cite{JohGIJ03}, hence we chose this scenario to be used in our
ENSO-DAO. Using the standard model of Section
\ref{sec-annualforcing}, this gives
\begin{equation}
\beta_{\rm global\ warming} = k^{-1} \sqrt{\frac{b}{k}} \
\frac{5^\circ \mbox{C}}{100 \mbox{years}} = 0.009
\end{equation}
as the value of $\beta$ in Eq.\ (\ref{eq-forcedDAO}).

\subsection{Fixed points, linear stability analysis and consequences}

The fixed point solutions are now determined by the equation:
\begin{equation}
(1-\alpha) \bar{T}^{(\beta)} - \left( \bar{T}^{(\beta)} \right)^3 +
\beta = 0
\end{equation}
For small $\beta$, there is hardly any change from the original
$\bar{T}$ fixed points, and hence the system will behave as the
basic DAO. As $\beta$ is increased, the unstable inner fixed point
$\bar{T}^{(\beta)}_0$ decreases from zero, whereas both the outer
solutions $\bar{T}^{(\beta)}_{\pm}$ increase. In cases when $\beta$
is large, and for large enough $\alpha$, only one fixed point
remains on the real axis (the other two are now complex).

\begin{figure}[tbhp]
\includegraphics[width=0.9\columnwidth]{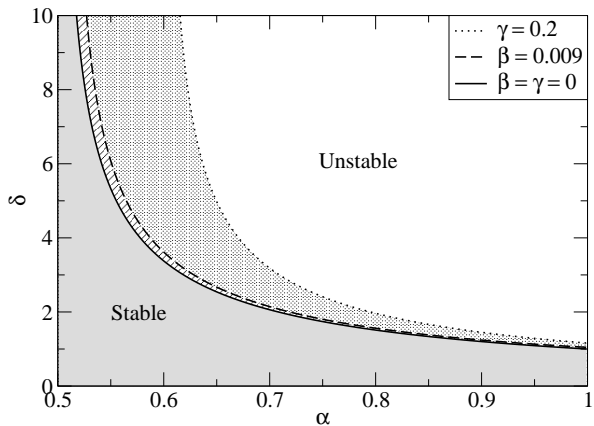}
\caption{
    The first neutral curve for the original DAO (solid black, as in Fig.\ \ref{fig-neutralcurves}),
    forced DAO (dashed black) and symmetrically coupled DAO (dotted
    black). Also shown is the stability region affected by global
    warming (hashed), and by coupling (dotted). Higher mode neutral curves are not shown.}
\label{fig-modneutralcurves}
\end{figure}

Instead of considering solutions starting near $\bar{T}_+ =
\sqrt{1-\alpha}$, we now start near $\bar{T}^{(\beta)}_+$. The
neutral curves can be determined via the same methods used to obtain
Eq.\ (\ref{eq-neutralcurve}), but with the fixed point
$\bar{T}^{(\beta)}_+$ now depending on $\beta$:
\begin{equation} \label{eq-betaneutralcurve}
\delta = \frac{2n \pi \mp \arccos \left[\frac{1 - 3
\left(\bar{T}^{(\beta)}_+\right)^2}{\alpha}\right] }{ \sqrt{
\alpha^2- \left[1 - 3 \left(\bar{T}^{(\beta)}_+\right)^2\right]^2
}}\ , \qquad n \in \mathbb{N}\cup\{0\}
\end{equation}
The first neutral curve for a $\beta_{\mathrm{global \ warming}}$
DAO system, compared to no heating, is plotted in Fig.\
\ref{fig-modneutralcurves}.

Consider a basic DAO system (above the solid black neutral curve in
Figure \ref{fig-neutralcurves}) that has an unstable fixed point
$\bar{T}^{(\beta=0)}_+$ with limit cycle oscillations. By
introducing $\beta>0$, the neutral curve is shifted to the right,
and hence the additional (hashed) stable region in Fig.\
\ref{fig-modneutralcurves} is created, changing the DAO dynamics. As
$\beta$ increases, the model fluctuates for longer about the
unstable $\bar{T}^{(\beta)}_+$ before flipping out to oscillate
around both fixed points. Eventually, the dashed neutral curve in
Fig.\ \ref{fig-modneutralcurves} moves to the right of ($\alpha,
\delta$) and $\bar{T}^{(\beta)}_+$ becomes attracting. This
stability bifurcation occurs at $\beta_{\mathrm{bif}}$, which can be
found for any given ($\alpha, \delta$) by solving Eq.\
(\ref{eq-betaneutralcurve}) for $\beta$.

If enough heating were applied to the DAO, the ENSO cycle would be
terminated, leaving an ocean at a constant temperature and with no El
Ni\~no events. However, for slow heating, such as $\beta=0.009$, there
is very little change to the time dependence of the DAO with respect to
SST anomaly amplitudes or their periodicity except for a small subset of
solutions. This would suggest that global warming will not affect ENSO,
and hence El Ni\~no events will still continue to occur.

\section{Stochastic Effects}
\label{sec-randomeffects}

Chaos inherent in real world weather systems plays an important part
in the irregularity of El Ni{\n}o events. Processes such as storms,
cyclones and ocean currents which traverse the Pacific play no part
in our ENSO model, but will influence the system for their duration
in the Ni{\n}o $3.4$ region. We therefore consider the addition of a
general stochastic term  \cite{Has76} $R(t)$ to the dimensional DAO
equation (\ref{eq-DAO}) giving:
\begin{equation} \label{eq-weatherDAO}
\frac{dT}{dt}=\ k T -\ b T^3 -\ {\cal A} T(t-\Delta) + \ R(t)
\end{equation}
$R(t)$ is a step function that takes fixed, normally distributed
(mean $0$, standard deviation $\lambda$), random values for each
monthly time-step. A (non-ENSO) event that lasts for longer than one
month is unphysical, and shorter-lived events will not have time to
cause any significant differences (their randomness is averaged
out).

Another way of viewing $R(t)$ is as a $\beta$ heating/cooling that
only lasts for one month. This will mean that the neutral curves
given by Eq. (\ref{eq-betaneutralcurve}) will shift right/left each
month, due to the fixed points $\bar{T}^{(R)}_\pm$ changing
position.

If $\lambda$ is small, then there is hardly any change to the basic
DAO. However, for large enough $\lambda$, the fixed points can
change their stabilities each month, inducing a great deal of random
behavior in the DAO. For example, the system may be stable and
tending towards $\bar{T}^{(R)}_+$ of that month, and then $R(t)$
changes to produce an unstable system that wants to oscillate. This
creates many different El Ni\~no events with differing amplitudes
and durations, as well as the possibility of times where the system
stays near zero.

\section{The full ENSO-DAO}
\label{sec-allinone}

\begin{figure}[tbhp]
\includegraphics[width=0.9\columnwidth]{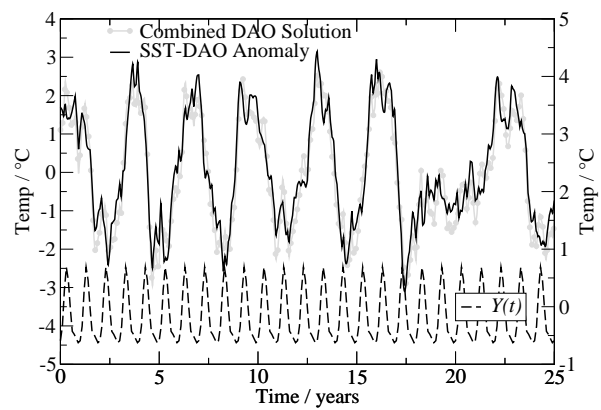}
\caption{
    Numerical solution of the combined DAO
    (\ref{eq-allinone}) with $\delta=3$ and $\alpha=0.7$ showing the annual cycle $Y(t)$ (dashed
    black, right axis), model solution $T$ (grey, left axis) and SST anomaly $T-Y(t)$ (solid
    black, left axis).}
\label{fig-allinone}
\end{figure}

We now combine of all of the effects considered above, namely the
annual cycle, global warming and stochastic effects. Therefore the DAO
equation (\ref{eq-DAO}) becomes
\begin{equation} \label{eq-allinone}
\frac{dT}{dt}=\ k T -\ b T^3 -\ {\cal A} T(t-\Delta) +\ Y'(t) +\
{\cal B} +\ R(t)
\end{equation}
Using the standard model of Section \ref{sec-annualforcing}, with
${\cal B}= 5^\circ\mbox{C}/100$years as the fully dimensional global
warming factor and $R(t)$ with $\lambda=5$, the DAO produces very
realistic pictures, such as Fig.\ \ref{fig-allinone}, that look
remarkably similar to Fig. \ref{fig-SSToscillations}.
Most obvious are the small fluctuations associated with the
stochastic effects. In addition, the regularity of the original DAO
oscillations is broken and sometimes a complete ENSO event has
vanished. E.g.\ the absence of the SST peaks at about $t\sim 20$ in
Fig.\ \ref{fig-allinone}. This happens when the onset of the ENSO
cycle event takes place at a time when the random forcing happens to
change the SST into the opposite direction. We emphasize that a
variety of solution curves similar to Fig.\ \ref{fig-allinone} can
be generated by taking other ($\alpha, \delta$) combinations,
initial SST values and stochastic forcing.


\section{Coupled DAOs}
\label{sec-coupledDAOs}

\begin{figure}[tbhp]
\includegraphics[width=0.9\columnwidth]{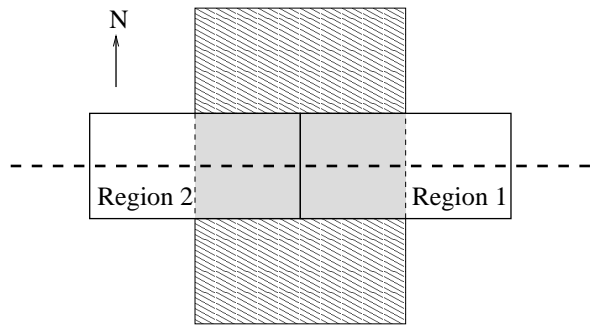}
\caption{
    Schematic of the Pacific showing the original ENSO-DAO region
    (grey), new coupled regions $1$ and $2$ and potential North--South
    regions (hashed). The equator is indicated as a horizontal dashed line.}
\label{fig-coupledregions}
\end{figure}

The basic DAO model (\ref{eq-DAO}) considers a single, but
representative region with strong atmospheric-ocean coupling in the
Pacific. Clearly, we might also want to consider situations in which
other regions of the Pacific are incorporated. These could be either
other regions along the equator, where we would expect differing
coupling strengths and delay times. Or they could be regions to the
North and South of the central DAO region in which no delay and only
weak coupling exists, such that these additional regions act as
temperature sinks. Possible arrangements are shown in Fig.
\ref{fig-coupledregions}. In this chapter, we will consider one such
prototype situation, namely two coupled regions along the equator.
\subsection{Two coupled DAOs along the equator}

Let us for simplicity assume that coupling $\alpha$ and delay
$\delta$ in the two regions are the same. This is a valid assumption
as long as the two adjacent regions are both in the strong-coupling
region such that the distance to the Western boundary is
approximately the same, with the same losses and reflection
properties.
If one region is warmer than the other, then the flow of heat energy
across their common boundary will act to heat one and cool the
other. This can be modeled by incorporating a $\gamma T_{i+1}$ term
in each DAO with $\gamma>0$ and $T_{i+1}$ the anomaly in the
adjacent DAO region. Then we have a system of $2$ coupled DAOs such
that
\begin{subequations}\label{eq-DAOcoupled}
\begin{eqnarray} \label{eq-region1}
\frac{dT_1}{dt}=\ T_1 -\ {T_1}^3 -\ \alpha T_1(t-\delta) + \gamma T_2 \\
\frac{dT_2}{dt}=\ T_2 -\ {T_2}^3 -\ \alpha T_2(t-\delta) + \gamma
T_1 \label{eq-region2}
\end{eqnarray}
\end{subequations}

The fixed point solutions can be found as before by taking
$\frac{dT_i}{dt}=0$ and substituting (\ref{eq-region1}) into
(\ref{eq-region2}). This yields a $9$th order polynomial in
$\bar{T}^{(\gamma)}_i$ with a mixture of real and complex roots
which vary as a function of $\gamma$. The solutions
$\bar{T}^{(\gamma)}_{i\pm} = \pm \sqrt{1-\alpha + \gamma}$ are the
same in both regions and are always real since $0<\alpha<1$, and
hence are of most interest.

Conducting a linear stability analysis about the fixed points
($\bar{T}^{(\gamma)}_{1+}$, $\bar{T}^{(\gamma)}_{2+}$) as in Section
\ref{sec-DAO-linearstability}, we find neutral curves given by
\begin{equation} \label{eq-gammaneutral}
\delta = \frac{2n \pi \mp \arccos\left(\frac{3 \alpha - 2
-2\gamma}{\alpha}\right) }{ \sqrt{ \alpha^2- \left(3 \alpha - 2
-2\gamma\right)^2 }}\ , \qquad n \in \mathbb{N} \cup \{0\}
\end{equation}
The first neutral curve is shown in Figure
\ref{fig-modneutralcurves} for an example with $\gamma=0.2$.

We see that similar to Section \ref{sec-globalwarming}, the neutral
curves move right with increasing $\gamma$, such that the fixed
point can bifurcate from unstable to stable. So, if there is strong
coupling between regions in the Pacific (large $\gamma$), then
independent oscillations cannot occur and the system settles down to
constant temperature anomalies. However, for sufficiently weak
coupling, ENSO-style oscillations are still permitted.

As before, we have also solved the Eq.\ (\ref{eq-DAOcoupled})
numerically. Depending on the initial SST anomalies, we find that we
can drive the regions into (i) stable, non-oscillatory solutions,
(ii) in-phase oscillations and even (iii) oscillations which are
out-of-phase by $\pi$. For brevity, we do not show the results here.

\subsection{Two regions with different ENSO coupling}

\begin{figure}[tbhp]
\begin{center}
\includegraphics[width=0.9\columnwidth]{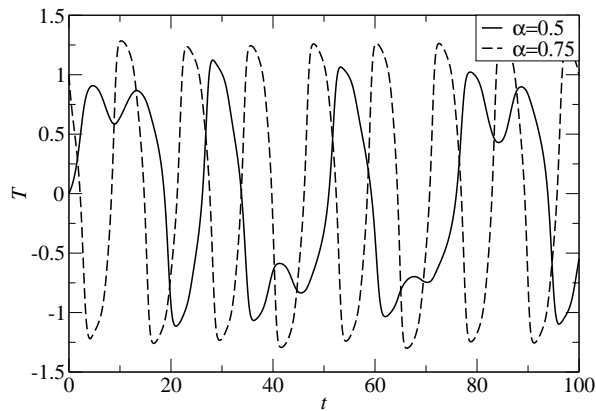}
\caption{
    Numerical solutions to the coupled DAO (\ref{eq-DAOcoupled}),
    with $\alpha_1=0.5$ in region $1$, $\alpha_2=0.75$ in region
    $2$, $\delta=4$ in both regions and $\gamma=0.1$.}
    \label{fig-differentalpha}
\end{center}
\end{figure}

Let us now consider briefly the situation in which $\alpha_i$ or $\delta_i$ are
different in the different regions.
If $T_2$ represents the anomaly in region $2$ of Fig.
\ref{fig-coupledregions} and $T_1$ the anomaly in region $1$, then
the wave transit time from $T_1$ to the western boundary would be
greater than that from $T_2$. Similarly, the strength of the
returning signal to $T_1$ would be less than that to $T_2$, as more
information would be lost due to the greater distance across which
information is transported. However, the waves propagating from
$T_1$ are not hidden from the problem whilst they are propagating
through $T_2$, suggesting that $\delta$ could be the same for both
regions, namely the propagation time from $T_2$ to the western
boundary and back. Similarly, the strength of the local coupling is
assumed to be less in $T_1$, suggesting that even though the
strength of the returning signal is less, relative to the local
coupling its affect would be the same as in $T_2$. Figure
\ref{fig-differentalpha} shows how the coupled model
(\ref{eq-DAOcoupled}) is affected by different values of $\alpha$.
In the Figure we see that $T_1$ shows some irregular behavior with
varying amplitudes. This behavior is more like that of El Ni\n o
than the simple DAO.

Other possible coupled DAO scenarios could include $3$ coupled regions, either
along the equator or aligned along a south-north axis. Our preliminary results
indicate that such models gives rise to irregular ENSO cycles as above. However,
the models have to be treated with proper care, since some solutions become
unphysical, e.g.\ indicating uninterrupted heating or cooling of the ocean.

\section{Conclusions}
\label{sec-conclusions}

The DAO is given by a simple one-dimensional differential equation,
and may nowadays be studied by readily available desk- or laptop
computing facilities. Yet it captures much of the observed richness
and variance of the ENSO phenomenon. We are therefore convinced that
it presents an ideal tool to introduce students to the topic.
Let us briefly recapitulate what issues we have been able to study in
the present paper using the DAO.
We first outline the linear stability analysis and the numerical solution of the
DAO that were the content of the first paper on the DAO \cite{SuaS88}. This then
serves to introduce the concept of shallow-water Kelvin and Rossby waves, as well
as to establish our notation. Using the corresponding wave speeds and the
geography of the Pacific, we can already show that for a large set of parameters,
the observed ENSO periodicity can be modeled by the DAO.

Next, the inclusion of the measured annual cycle demonstrates that the
observed ENSO-DAO periodicity may in fact be heavily influenced by the
Earth's period around the Sun. The occurrence of an ENSO event is very
much subject to the right season and thus the original DAO period need
not be identical to the observed ENSO-DAO period. We stress that the
modeling of the annual cycle within the DAO is quite easy and should
not be beyond any student who has mastered the original DAO.

We then extended the DAO to include a constant heating
--- global warming --- and found that current predictions of
$5^\circ$C rise over $100$ years are unlikely to influence the
ENSO-DAO dynamics. We also do not find any large increase in either
occurrence or magnitude of ENSO events. This seems to be at variance
with current predictions based on CGCMs \cite{VogL98,TimOBE99} and
might just indicate the limitations of the ENSO-DAO model for true,
quantitative predictions \footnote{We remark
  that the issue of the interplay between global warming and El Ni\~{n}o
  is rather controversial and various differing opinions are offered
  \cite{Sin99}.}.

Putting it all together, we then included a certain level of
stochastic forcing alongside the annual cycle, and found that the
ENSO-DAO dynamics now become very much like the observed ENSO
dynamics, showing a large variation in oscillation periods and
amplitudes. At least on a qualitative level
--- the range of the SST oscillations is in fact even quantitatively
similar
--- which shows the strength of the ENSO-DAO.

Finally, we briefly discussed some extensions of the ENSO-DAO model,
such as coupled regions. The example of the $2$-coupled-regions
model shows that the extended DAOs share some irregularities with
the observed ENSO phenomenon already, even in the absence of annual
or stochastic forcing.
Further investigations of ENSO-DAO might include an analysis of the
statistical properties of the measured ENSO cycles as in Fig.\
\ref{fig-allinone} and comparisons with observed data, a more
thorough analysis of the possible changes in periodicity and
strengths of ENSO-DAO oscillations due to human influences, a
mathematical analysis of the DAO for integer periods or even an
application to the Atlantic.

In conclusion, we see that the DAO --- while being a research tool
in its own right \cite{WhiTBD03} --- can be investigated by modest analytical and
computational resources. We think that it will be an ideal topic for
all students, teachers and lecturers interested in the exciting
dynamics of the El Ni\~no phenomenon and hope that the present paper
will help them in this quest.

\acknowledgments
The present paper has grown from a final year student project at the
University of Warwick supervised by one of us (RAR). We thankfully
acknowledge discussions and careful reading of the manuscript by G.\
King and G.\ Rowlands. We thank C.\ Sohrmann for help with Fig.\
\ref{fig-DAOperiods}.

\appendix

\section{Shallow water equations and waves}
\label{sec-shallowwaterwaves}

\subsection{Two-fluid model and shallow water equations} Oceans are
warmer near the surface than they are in the deep, and due to
thermal expansion this warm layer will have a lower density. Thus a
simple model for the ocean structure is a two-fluid model with
density $\rho_1$ in the warm surface layer and density
$\rho_2>\rho_1$ below in the cold abyss. The dividing zone is called
the {\em thermocline}.
At its widest, the Pacific is $17,700$km across from the Malay
Peninsula to Panama whilst only having an average depth of $4,280$m.
So the Pacific is roughly $3$ orders of magnitude wider than it is
deep, and so we are justified in rewriting the original
Navier-Stokes equations using the {\em shallow-water approximation}
\cite{Gil82,Hol04}. These are given as
\begin{subequations}\label{eq-shallowwater}
\begin{eqnarray} \label{eq-continuity}
\pderiv{\eta}{t}+H \left(\pderiv{u}{x}+\pderiv{v}{y} \right) & = &0 \\
\label{eq-shallowx}
\pderiv{u}{t}+u\pderiv{u}{x}+v\pderiv{u}{y}-fv+g'\pderiv{\eta}{x}& = &0 \\
\label{eq-shallowy}
\pderiv{v}{t}+u\pderiv{v}{x}+v\pderiv{v}{y}+fu+g'\pderiv{\eta}{y}& =
&0
\end{eqnarray}
\end{subequations}
in the absence of external forces. Here $(u,v,w)$ is the
displacement velocity in the Cartesian frame $(x,y,z)$ and $\eta
(x,y,t)$ is a small perturbation to the constant mean thermocline
depth $H$. Furthermore, $f$ is the Coriolis parameter and
$g'=\frac{\rho_2-\rho_1}{\rho_1} g$ is the effective gravitational
acceleration.
Equations (\ref{eq-shallowwater}) permit two types of thermocline wave solutions.
The first are inertia-gravity {\em Kelvin} waves. The second type
are planetary {\em Rossby} waves.

\subsection{Kelvin waves}

Equatorially trapped Kelvin waves have no meridional ($y$) velocity
fluctuations such that the shallow water equations
(\ref{eq-shallowwater}) can be simplified to
\begin{subequations}
\begin{equation}\label{eq-kelvinx}
\pd{u}{t}-g'H\pd{u}{x}=0
\end{equation}
\begin{equation}\label{eq-kelviny}
f\pderiv{u}{t}-g'H\frac{\partial^2 u}{\partial x \partial y}=0
\end{equation}
\end{subequations}
Equation (\ref{eq-kelvinx}) implies that solutions have to be of the
form $u=E(y)F(x \pm ct)$ with wave speed $c=\sqrt{g'H}$. In
addition, Eq.\ (\ref{eq-kelviny}) prohibits westward propagation
($F(x+ct)$) since those solutions become unbounded at large $y$.
Therefore only eastward, equatorially trapped ($v=0$), Kelvin waves
are possible.
These non-dispersive thermocline gravity waves propagate eastward
with speed $c\approx 1.4$ms$^{-1}$, when assuming a height $H=100$m
and $(\rho_2-\rho_1)/\rho_1=0.002$ \cite{Phi90} for the two-fluid
model \footnote{If a single-fluid model of the ocean with depth
$H=4000$m had been assumed throughout, then $c = \sqrt{gH}
\approx200$ms$^{-1}$. This Tsunami-like speed is clearly absurd for
the purposes of this project, hence the need for the more
sophisticated two-fluid model.}.

\subsection{Rossby waves}

Rossby waves owe their existence to the variation of the Coriolis
parameter with latitude.
Close to the equator the Coriolis parameter is given by $f=2\Omega
y/r_{\rm Earth}$ where $r_{\rm Earth}$ is the radius of the Earth
and $\Omega$ the Earth's angular velocity. If we consider a
linearized version of the shallow-water equations and seek solutions
for latitudinal waves of the form $v=V(y)\exp{\imath(kx- \omega
t)}$, with $\omega>0$ and the sign of $k$ determining the direction
of zonal ($x$) phase propagation, we obtain their wave equation as
\begin{equation} \label{eq-rossby}
\pd{V}{y}+\frac{\beta^2}{g'H}\left(Y^2-y^2\right)V=0
\end{equation}
with
\begin{equation}\label{eq-waveguide}
    Y^2 = \frac{g'H}{\beta^2} \left( \frac{\omega^2}{g'H} - k^2
- \frac{\beta k}{\omega} \right)
\end{equation}
and $\beta=2\Omega/r_{\rm Earth}$. From (\ref{eq-rossby}) we see
that solutions are wavelike within an {\em equatorial waveguide} of
width $2Y$, and decay exponentially for latitudes greater than
$|Y|$.

If we now assume that the latitudinal dependance of $V$ is of plane-wave type
$\sim \exp(\imath ny)$, then for these equatorially trapped waves we have $n \in
\mathbb{N}$ as the only allowed standing wave solutions.
A dispersion relation can be extracted from (\ref{eq-rossby}) and at
low frequencies
\begin{equation}\label{eq-dispersion}
    \omega=\frac{-\beta k}{k^2+\frac{2n+1}{\lambda^2}} \ \ , \ \quad n \in \mathbb{N}
\end{equation}
with wavelength $\lambda=\sqrt{c/\beta}$. Since $k$ must therefore be
negative, all Rossby waves have westward phase propagation. The slow,
short, dispersive waves have eastward group velocities and the fast,
long, non-dispersive waves have westward group velocities of
$c/(2n+1)$. Only the long waves are important in the oceanic adjustment
to changes in surface winds --- a process that is crucial in
ocean-atmosphere coupled systems --- so only westward propagating
Rossby waves will usually be considered.

The fastest Rossby wave ($n=1$) travels at one third the speed of a
Kelvin wave, e.g.\ $\approx 0.47$ms$^{-1}$ in the Pacific. These are
the non-dispersive, equatorially trapped, westward propagating
Rossby waves considered in the DAO model in Section \ref{sec-DAO}.

\section{Numerical Routines}
\label{sec-numericalroutines}

We include an abbreviated, but fully functional {\sl Mathematica
5.2} code for the solution of the annually forced DAO with global
warming and monthly random weather as studied in Section
\ref{sec-allinone}. Physical units used in the code are given in
years for time and Kelvin for temperature throughout.

\footnotesize
\begin{verbatim}
<<NDelayDSolve.m

(* Define the variables to be used and calculate
   the required scalings *)

year=365.24; delay=349/year; alpha=0.7; delta=3;
beta=0.05; k=delta/delay; time=25;

(* Solve the basic dimensionless DAO, determine its
   maxima and therefore the value of b, the
   temperature scaling *)

Tfp=Sqrt[1-alpha];
orig= NDelayDSolve[
       {T'[t] == T[t]-T[t]^3-alpha*T[t-delta]},
       {T->(Tfp+0.05&)},
       {t,0,25*delta},
       MaxRecursion->50
      ];
values=Table[T[t]/.orig,{t,0,25*delta,0.1}];
Tmax=Max[values];
b=k*(Tmax/2)^2;

(* Read the real data from file, and construct
   annual cycle function *)

data=ReadList["ave_sea_temp.txt"];
yearly = Interpolation[data-27.1,
          PeriodicInterpolation->True];
Y[t_]:=yearly[ 12*t +1 ]

(* Create a Gaussian Distributed random variable
   (mean zero, standard deviation lambda) which
   changes every month, representing the effects
   of stochastic processes on the DAO *)

lambda=5;
Needs["Statistics'ContinuousDistributions'"];
stochastic=Table[Random[
       NormalDistribution[0,lambda]],{time*12+1}
       ];

R[t_]:= stochastic[[IntegerPart[12*t]+1]];
Plot[R[t],{t,0,time}]

(* Solve the Combined DAO Equation *)

Tf = Extract[T /.
  Solve[k*T - b*T^3 - k*alpha*T + beta == 0, T],
 3];

combo =
 NDelayDSolve[
  {(T')[t] == k*T[t] - b*T[t]^3
   - k*alpha*T[t - delta/k]
   + beta + R[t] + (Y')[t]},
   {T->((Tf + 0.15 &))
  },
  {t, 0, time},
  MaxRecursion -> 100, MaxSteps -> Infinity
 ];

(* Plot the Combined DAO Solution, the Annual Cycle,
   and the SST-DAO Anomaly *)

Plot[
 {Evaluate[T[t]/.combo],Y[t],Evaluate[T[t]/.combo]
   -Y[t]},{t,0,time},
  AxesLabel->{"t/years","T/K"},
  PlotStyle->{{RGBColor[0,1,0]},
              {RGBColor[0,0,0]},
              {RGBColor[1,0,0]}}
];

(* Calculate the period of the anomaly
   oscillations using Fourier Transform method *)

data=Abs[Fourier[Table[(T[t]/.combo)-Y[t],
    {t,0,time,0.1}]]];
Do[If[data[[k]]==Max[data],ans=Return[k]],
    {k,1,time,1}];
period=N[time/(ans-1)]

\end{verbatim}
\normalsize


\end{document}